\documentclass[12pt]{iopart}

\usepackage{iopams}

\usepackage{graphicx}

\begin{document}

\title[Degradation and destruction of historical blue-green glass beads]{Degradation and destruction of historical blue-green glass beads: A study by microspectroscopy of light transmission }

\author{Tatyana V Yuryeva$^1$\footnote[8]{http://www.gosniir.ru/about/gosniir-structure.aspx
} and Vladimir A Yuryev$^2$\footnote[6]{http://www.gpi.ru/eng/staff\_s.php?eng=1\&id=125}}

\

\address{$^1$ The State Research Institute for Restoration of the Ministry of Culture of Russian Federation, bldg 1, 44 Gastello Street,  Moscow, 107114, Russia}
\ead{tvyur@kapella.gpi.ru}

\

\address{$^2$ A.\,M.\,Prokhorov General Physics Institute of the Russian Academy of Sciences, 38 Vavilov Street, Moscow, 119991, Russia}
\ead{vyuryev@kapella.gpi.ru}

\begin{abstract}
Blue-green historical beads are sometimes referred to as instable ones because of their degradability. At present, the cause of the phenomenon of deterioration of the blue-green beads is unknown. We explore internal microstucture of degrading blue-green historical beads and its evolution in the process of bead deterioration. Investigating transmittance and scattering spectra of visible and near infrared light we observe formation of microscopic internal inhomogeneities with the sizes less than 150\,nm in the glass bulk and growth of their density with increase in degree of bead degradation. By means of laser scanning microscopy we also observe numerous microinclusions and microcracks  on the cleavage surface of a partially degraded bead. We discuss possible physical factors resulting in destruction of the blue-green beads.
 \end{abstract}

\vspace{2pc}

\submitto{\JOA}

\maketitle

\section{\label{sec:intro}Introduction. Background and Problem Statement}

\begin{figure}[t]
\centering
\includegraphics[scale=.8]{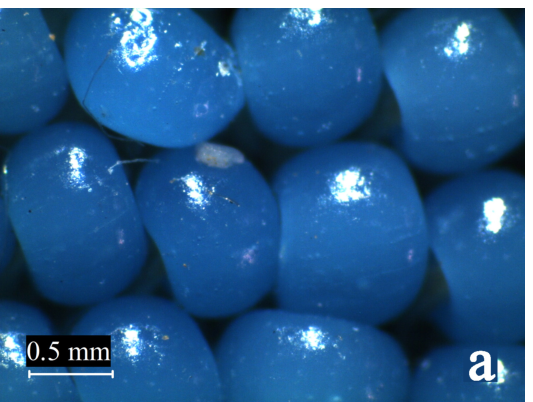}
\includegraphics[scale=.8]{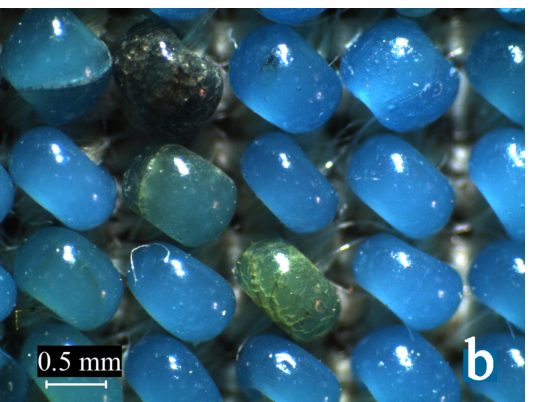}
\includegraphics[scale=.8]{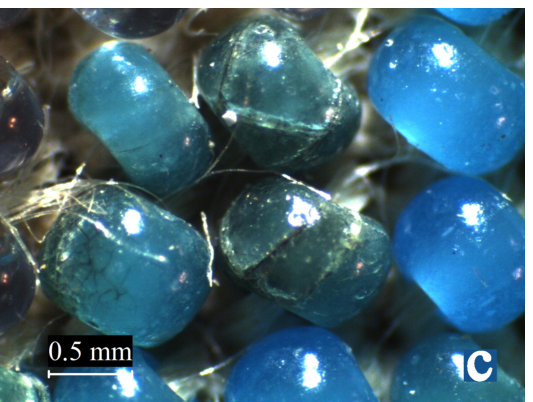}
\includegraphics[scale=.8]{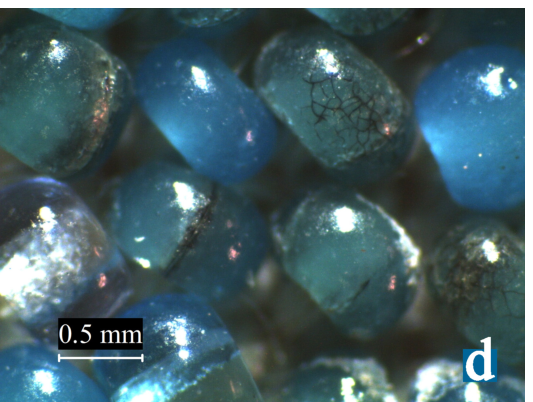}
\includegraphics[scale=.8]{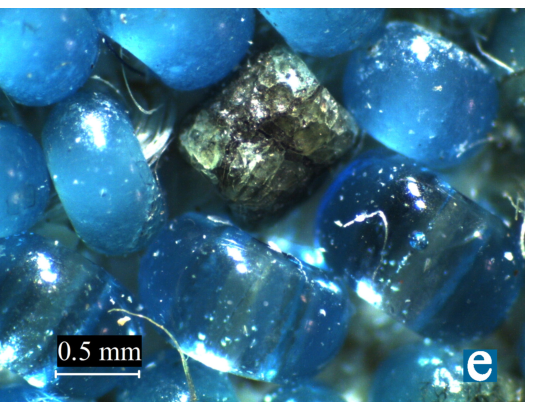}
\caption{\label{fig:beadwork}(Color online) 
Macro photographs showing historical beadwork of the blue-green beads (from early to mid 19th century; from the collection of the State Museum of A.\,S.\,Pushkin, Moscow, Russia):
(a) rows of intact beads (the undamaged historical beads are turbid and look luminous from the inside when scatter incident light);
(b) to (d) degraded beads at different phases of corrosion among the undamaged ones; 
(e)  a strongly degraded bead among well preserved historical ones (the opaque (turbid) ones in the upper left corner) and modern beads replacing the lost historical ones (the transparent ones in the lower right corner; the modern beads are bigger than the historical ones).
 }
\end{figure}  

\begin{figure*}[t]
\centering
\includegraphics[scale=1]{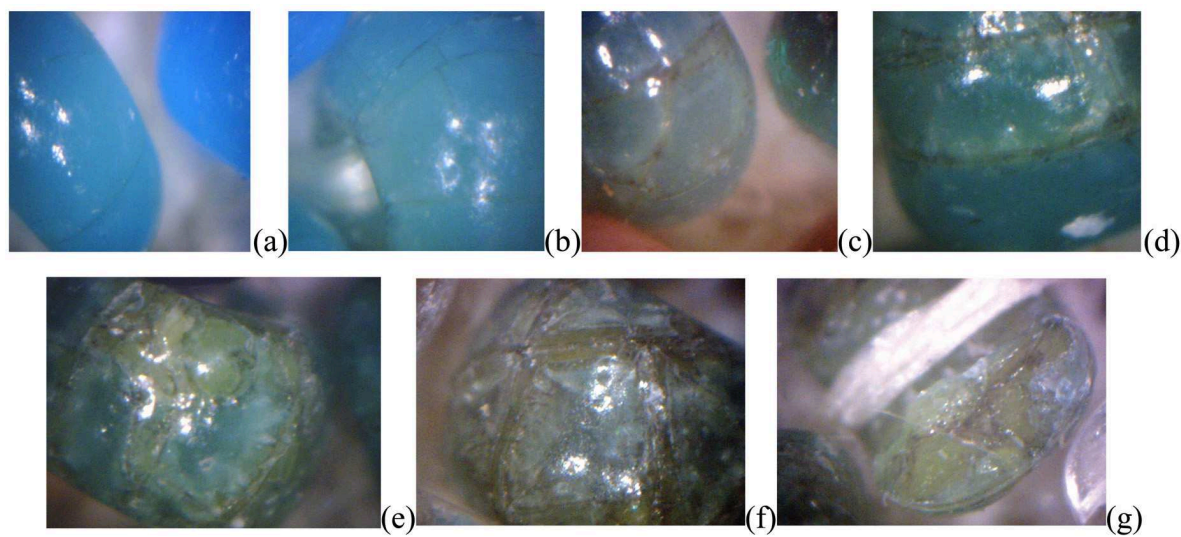}
\caption{\label{fig:phases} (Color online) 
Microphotographs demonstrating different phases of the blue-green beads destruction (the beads are sewn onto a cloth):
(a) cracks appear in blue beads; 
(b) cracking increases, the colour starts to change into the greenish tone; 
(c, d) further change of the colour, cracks change their colour in the greatest degree and become brown-green; 
(e)  discolouration of beads begins, the surface corrosion develops; 
(f) beads are completely faded; 
(g) beads fragment (the demonstrated bead segment remains sewn to a fabric with a white silk thread).
 }
\end{figure*}

In many museums throughout the world, historical art articles of beads made in different techniques and in different epochs are kept. In the years since their manufacture, surfaces of some beads have corroded, cracks have appeared in them, colours of some kinds of beads have changed, some beads have broken up into parts, i.\,e. beads degrade and crumble  \cite{Canadian_beads_Deterioration,Fertikov,Report_GosNIIR}.\footnote{
There exist a wide bibliography on structure, composition, corrosion, deterioration and alteration of archaeological and historical glasses and glassware \cite{Glasses_Bibliometry}. For details on degradation mechanisms of these kinds of glass and methods of their analysis, the readers can address, e.\,g., to  Refs. \cite{Corrosion_Sci, Archaeological_glass, Degradation_processes, Late_Roman, Methods}. However, the number of publications on a narrower issue of degradation and corrosion of historical or artistic articles of glass beadwork kept in museums is much less \cite{Glasses_Bibliometry}; some important information on this topic, devoted to ethnographic beads, can be found in the book \cite{Canadian_beads_Deterioration}.
Researches devoted to degradation of secular and church articles of beads kept in museums, which are less subjected to aggressive agents, are virtually   unavailable in the literature.
} 
It is known from the practice of museum keeping of articles made of beads that blue-green beads are more subjected to strong deterioration than ones of other colours \cite{Fertikov,Report_GosNIIR}. At the same time, both undamaged beads of this colour and beads at various phases of deterioration can be simultaneously present, e.\,g., in articles made from early to mid 19th century embroidered with beads (Fig.\,\ref{fig:beadwork}; the main stages of deterioration of the blue-green beads of the first half of the 19th century are illustrated by Fig.\,\ref{fig:phases}). Moreover, deteriorated  museum blue-green beads of the first half of the 19th century superficially resemble ones of the same colour found during the archaeological dig of a mound near Kholmy village in Russia \cite{Arch-Mos-Reg} (the mound is dated 17th to 18th century \cite{Kholmy_kurgan}) which are dramatically damaged.

We have found recently that aging of beads goes through a number of phases common to all the studied samples \cite{Report_GosNIIR}. They are as follows (Fig.\,\ref{fig:phases}): First cracks appear in the blue beads; then bead cracking increases and the bead colour starts to change into the greenish tone; after that the bead colour goes on changing, cracks change their colour in the greatest degree and become brown-green; then discolouration of beads begins, the surface corrosion develops; at the last stage, beads become completely faded and finally they fragment.

However, causes of degradation of blue-green beads have not been clarified thus far that prevents a development of restoration techniques  and conservation conditions of articles made of such beads.
In this connection, ascertainment  of these causes is believed to be of special importance.

Unfortunately, degrading blue-green beads are usually explored by means of integral analytical techniques and their microstructure is investigated very rarely \cite{Methods}. It is commonly adopted that reasons of bead deterioration are purely chemical, i.\,e. only chemical reactions going on a bead surface are responsible for its degradation: a high concentration of potassium oxide ($> 20$\,wt.\%) gives rise to formation of a considerable amount of potassium silicate and high rate of  hydrolysis of potassium silicate on the surface result in corrosion and destruction of this kind of beads \cite{Fertikov}. In archaeological beads found at Kholmy village \cite{Arch-Mos-Reg}, potassium oxide is contained in much less quantities ($< 7$\,wt.\%). However, they have strongly degraded in spite of much less content of K. 

Additionally, adjacent beads in a historical beaded article are often seen to be at radically different phases of corrosion---heavily degraded beads often adjoin intact ones (Fig.\,\ref{fig:beadwork})---while they are obviously very similar and seem to have close chemical composition including potassium content.
 
These facts make us assume that some other processes may be responsible for the blue-green bead destruction and try to find out what can be a cause of this ``selective'' corrosion of similar beads.

In this article, we explore an internal microstucture of the degrading blue-green historical beads and its evolution in the process of bead deterioration. Investigating transmittance and scattering spectra of visible and near infrared light we observe formation of microscopic internal inhomogeneities in the glass bulk and growth of their density with increasing degree of bead degradation.
We also discuss possible physical factors resulting in destruction of the blue-green beads.

\begin{figure*}[t]
\centering
\includegraphics[scale=1.2]{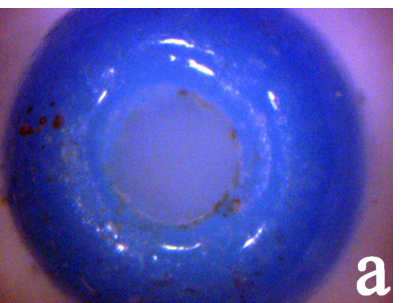}
\includegraphics[scale=1.2]{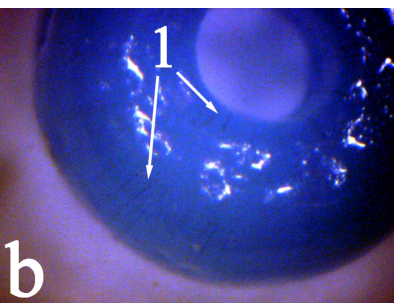}\\
\includegraphics[scale=1.2]{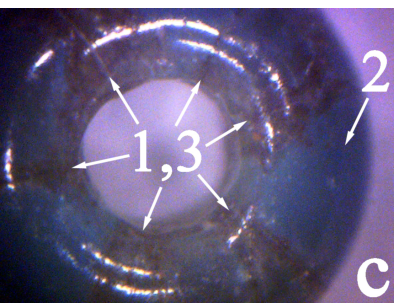}
\includegraphics[scale=1.2]{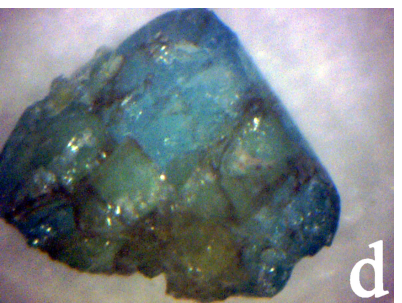}
\includegraphics[scale=1.2]{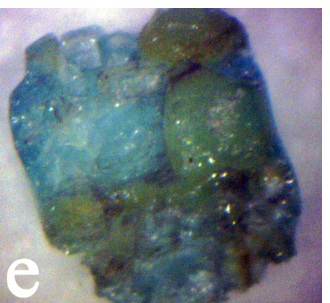}
\caption{\label{fig:photos} (Color online) 
Microphotographs of the investigated blue-green beads representing different stages of the degradation:
(a) an intact blue bead (sample 1);
(b) a cracked bead at the initial phase of destruction (sample 2);
(c) a strongly degraded bead with blue and brown-green (heavily cracked) segments (sample 3);
(d),  
(e) pieces of a fragmented bead, a conglomerate of differently coloured granules is seen to compose the parts of the destroyed bead (samples 4a and 4b);
figure 1 in panels (b) and (c) shows cracks in the beads;
figures 2 and 3 in panel (c) indicate blue and brown-green segments of the bead, respectively; the scale mark is given in Fig.\,\ref{fig:beadwork}. 
 }
\end{figure*}

\section{\label{sec:exp}Experimental Details}

\subsection{\label{sec:samples}Samples}

Photographs of the samples investigated by the light transmission spectroscopy  are shown in Fig.\,\ref{fig:photos}. We have chosen representative set of samples corresponding to major phases of the blue-green beads destruction. Sample 1 is an example of the intact blue bead; it is lustrous opaque-blue with no visible traces of cracking.
Sample 2 represents a bead at the very initial phase of destruction; it is seen to be cracked but still opaque-blue without greenish tinge, the cracks and regions around them do not change their colour either.
Sample 3 is a strongly degraded bead with blue and brown-green segments; the blue domains are of greenish tint; the brown-green segments are seen to drastically change their colour and be strongly cracked.  
Samples 4a and 4b are parts of a destroyed bead; they are composed by differently coloured grains, often also cracked, separated by sharp boundaries.

\subsection{\label{sec:equipment}Equipment and techniques}

For experiments on spectroscopy of light transmitted by beads, a home-made microspectrometer was assembled from the Avantes AvaSpec-2048 fiberoptic spectrometer attached to the LOMO Biolam~70-P13  transmitted light stereo microscope.  A tungsten lamp with a frosted glass bulb was used for sample illumination. The light transmission spectra were obtained at magnification of 600$\times$ in the range from violet to near-infrared light. The procedure of obtaining characteristic spectra was as follows: Several transmittance spectra, $Tr(\lambda)$, were measured at similar points of each sample (or at similar points of each specific domain of a sample) with the required integration time and averaging over an appropriate number of scans, then after comparison the obtained spectra were averaged over every set of analogous points for further analysis  of features typical for different stages of the bead degradation.  

Additionally, a Carl Zeiss LSM-710 laser scanning microscope (LSM) was applied to obtain detailed images of cleavage surfaces of the destroyed beads. An Eleran Renom FV X-ray fluorescence (XRF) spectrometer was used for the elemental analysis of beads.

\section{\label{sec:results}Results and Discussion}

\subsection{\label{sec:XRF}XRF Spectral Analysis}

\begin{table}
\caption{\label{tab:XRF-Analysis}Typical data of the XRF chemical analysis of two blue-green beads. The table presents values of the XRF line intensity ({$I_{\rm xrf}$}) and a relative concentration ({$C_{\rm rel}$}) for revealed chemical elements. It also gives weight fractions of corresponding oxides ($F_{\rm ox}$) in the glass.} 

\begin{indented}
\lineup
\item[]\begin{tabular}{@{}llcccccc}
\br  
&&&&&\centre{3}{A bead with blue and}\\
&&\centre{3}{A blue bead}&
\centre{3}{brown-green segments}\\
\ns  
&&\crule{3}&
\crule{3}\\                          
Chemical&&
\centre{1}{$I_{\rm xrf}$}&
\centre{1}{$C_{\rm rel}$} &$F_{\rm ox}$&
\centre{1}{$I_{\rm xrf}$}&
\centre{1}{$C_{\rm rel}$}&$F_{\rm ox}$\cr 
element &Oxide&
\centre{1}{(counts)}&
\centre{1}{(wt.\%)}&(wt.\%)&
\centre{1}{(counts)}&
\centre{1}{(wt.\%)}&(wt.\%)\cr 
\mr
Si&SiO$_2$&118.76&63.38&75.95&\079.51&52.35&67.02\cr
Cu&CuO&303.54&\05.00&\03.51&150.21&\03.09&\02.31\cr
K&K$_2$O&225.71&10.97&\07.40&267.80&14.20&10.24\cr
Pb&PbO&\058.96&16.53&\09.97&\064.40&25.20&16.25\cr
Ca&CaO&\033.40&\01.93&\01.51&\033.24&\01.49&\01.25\cr
Fe&Fe$_2$O$_3$&\015.86&\00.36&\00.29&\012.32&\00.36&\00.31\cr
As&As$_2$O$_3$&347.80&\01.31&\00.97&405.58&\01.71&\01.35\cr
Sb&Sb$_2$O$_5$&\0\02.58&\00.52&\00.39&\0\05.44&\01.60&\01.27\cr
\br
\end{tabular}
\end{indented}
\end{table}

To characterize glasses of which the studied beads are made chemical analysis of some blue-green beads was  accomplished by means of XRF spectroscopy. Typical results of such analysis obtained from a blue bead and from a bead with blue and brown-green segments are given in Table\,\ref{tab:XRF-Analysis}. The table presents values of the XRF line intensity ({$I_{\rm xrf}$}) and a relative concentration ({$C_{\rm rel}$}) for detected chemical elements; it also gives weight fractions of corresponding oxides ($F_{\rm ox}$) in the glass which are usually utilized in glass manufacturing. $C_{\rm rel}$ are normalized in such a way to give a sum total  of 100\,wt.\%. The values of $F_{\rm ox}$ are calculated from $C_{\rm rel}$ and also normalized to give a sum total  of 100\,wt.\%.

The beads are seen to be made of lead-potassium glass with rather high content of copper; appreciable quantities of calcium, arsenic, antimony and iron have also been detected in the glass. CuO dyes the glass into blue or slightly greenish colour. We suppose that Sb is present in glass in the form of Sb$_2$O$_5$ (if the prevailing oxide is Sb$_2$O$_3$ the estimates of concentrations of the oxides remain practically unchanged within the limits of the experimental error). Antimony likely enters glass from charcoal ash which was often used as a source of potassium (potash) although Sb (Sb$_2$O$_3$) might also be  added to the glass charge intentionally, together with As$_2$O$_3$, as opacifyer (both antimony and arsenic was used for opacifying beads in the early 19th century \cite{EuroBeads}) or as oxidant to ensure prevailing of the iron impurity in the  Fe$^{3+}$ state (Fe$_2$O$_3$) rather than Fe$^{2+}$ (FeO) since the latter in the mixture with ferric oxide colours glass in green.

We should mention also that the intact blue bead contains by several times less antimony than the degraded one while the fractions of potassium oxide in the samples differ only by about 1.4 times.

\subsection{\label{sec:transmittance}Light Transmittance}

\begin{figure}[t]
\centering
\includegraphics[scale=1]{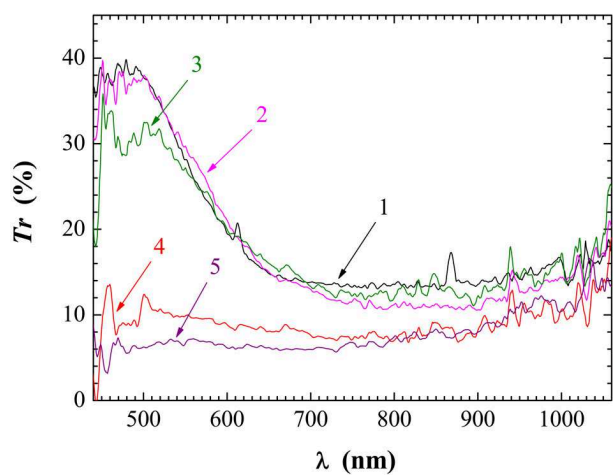}

\caption{\label{fig:tr} (Color online) 
Transmittance ($Tr$) spectra of the blue-green beads shown in Fig.\,\ref{fig:photos} derived as a result of averaging of a number of spectra obtained from similar points on each bead or on a specific segment of the bead at magnification of 600: 
(1) an intact blue bead (sample 1, Fig.\,\ref{fig:photos}\,a),
(2) a cracked bead at initial phase of destruction (sample 2, Fig.\,\ref{fig:photos}\,b),
(3)  a strongly degraded bead (sample 3, blue segments, Fig.\,\ref{fig:photos}\,c, domain 2), 
(4) a strongly degraded bead (sample 3, brown-green segments, Fig.\,\ref{fig:photos}\,c, domain 3)
and
(5)  a disrupted bead (samples 4a and 4b, Fig.\,\ref{fig:photos}\,d,\,e).
}
\end{figure}

The resultant transmittance spectra of the examined beads obtained following the procedure described above (Sec.\,\ref{sec:equipment}) are presented in Fig.\,\ref{fig:tr}. The spectra of the samples 1 and 2 (curves 1 and 2) are very close in the whole spectral range. This means that arising cracks makes no contribution to  absorption or scattering of light. 

The spectra of the blue region of the sample 3 (curve 3) also nearly coincide with those of the intact sample throughout the studied spectral range except for the interval from green to blue light  ($\lambda < 550$\,nm) where $Tr(\lambda)$ of the sample 3 first slightly and then ($\lambda < 500$\,nm) significantly  decreases in comparison with $Tr(\lambda)$ of the sample 1. This behavior of $Tr(\lambda)$ reflects the greenish colour of these domains; it may be caused by  growth of either absorption or scattering or by simultaneous increase of both absorption and scattering.

A dramatic reduction of transmittance is observed in the spectra obtained from the brown-green segments of the sample 3 (curve 4) and in the spectra of the sample 4 (curve 5). This effect also may result from significantly increasing absorption and/or scattering of light in these samples.

\subsection{\label{sec:scattering}Rayleigh Scattering of Light}

\begin{figure}[t]
\centering
\includegraphics[scale=1]{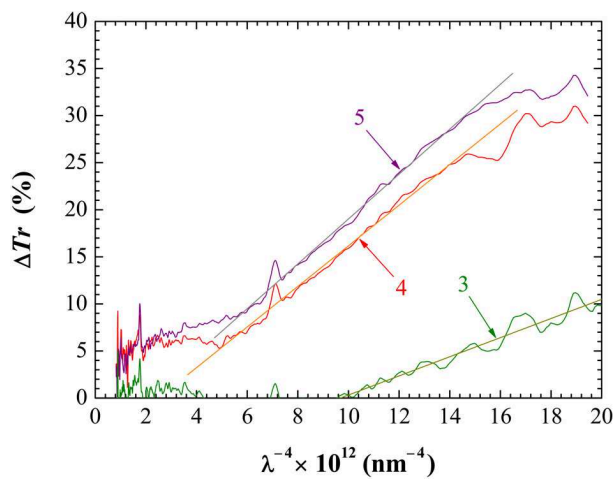}
\caption{\label{fig:scatter}
A result of subtraction of the spectra (3) to (5) from the spectrum (1) ($\Delta Tr$, see Fig.\,\ref{fig:tr}) plotted as a function of $\lambda^{-4}$ and linear fits of fragments of the graphs in the range of the visible light; designations are the same as in Fig.\,\ref{fig:tr}.
}
\end{figure}

We can easily discriminate between reduction of $Tr$ caused by growing absorption  and that caused by increasing scattering of light, if one of this factors prevails, by subtracting $Tr_{i>1}(\lambda)$ from $Tr_{i=1}(\lambda)$ ($\Delta Tr(\lambda)= Tr_{i=1}(\lambda) - Tr_{i>1}(\lambda)$), where $i$ is a number of the sample, and plotting the result $\Delta Tr$ as a function of $1/{\lambda}^4$. If the differential spectra obey the Rayleigh law, i.\,e. if $\Delta Tr\propto \lambda^{-4}$, the light scattering dominates; otherwise the impact of light absorption is significant.

Fig.\,\ref{fig:scatter} demonstrates the $\Delta Tr(\lambda^{-4})$ spectra for the samples 3 and 4 and linear fits of their fragments in the range of the visible light. All the presented spectra, even the spectrum obtained in the blue-green segments of the sample 3,  are seen to strictly obey the Rayleigh law, so we can conclude that the observed changes in $Tr(\lambda)$ result from increasing light scattering in the bulk of the degrading beads.

\subsection{\label{sec:cracks}Microinclusions, Microcracks and Internal Stress}

From the $\Delta Tr(\lambda^{-4})$ spectra given in Fig.\,\ref{fig:scatter}, we can estimate the characteristic dimensions of the scatterers as $a \ll \lambda_{\rm min}/\pi \sim 150$~nm. Hence, we can make an assumption about possible nature of the inhomogeneities scattering light. They are likely some microscopic inclusions in the glass of the beads or tiny cracks. Their density is obviously much higher in the vicinity of the observed large cracks of the sample 3 and in granules or at grain boundaries of the sample 4 than in the relatively undamaged blue-green domains of the sample 3.

\begin{figure*}[t]
\centering
\includegraphics[scale=.7]{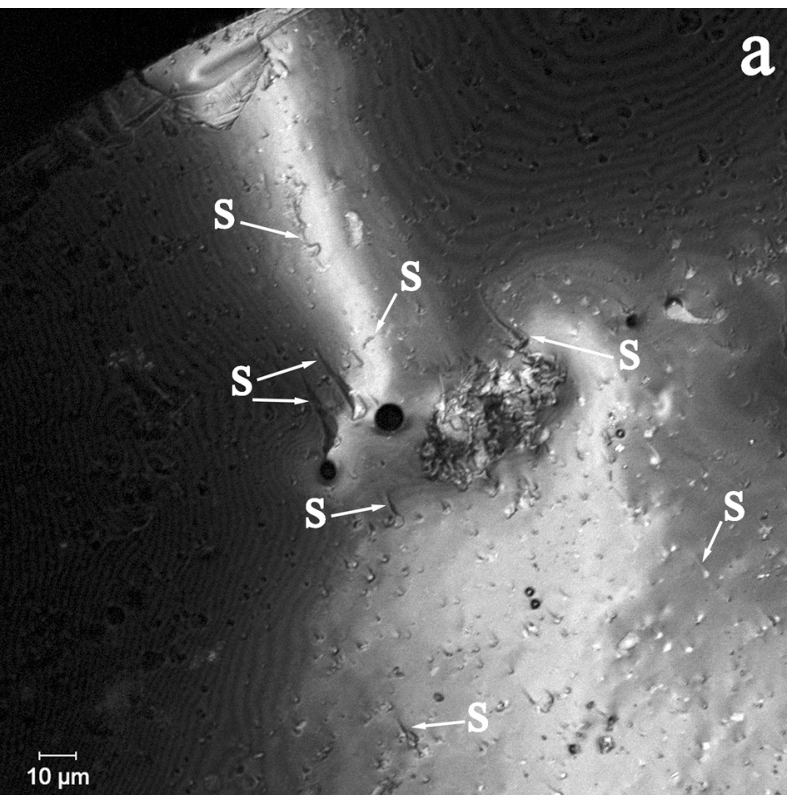}
\includegraphics[scale=.7]{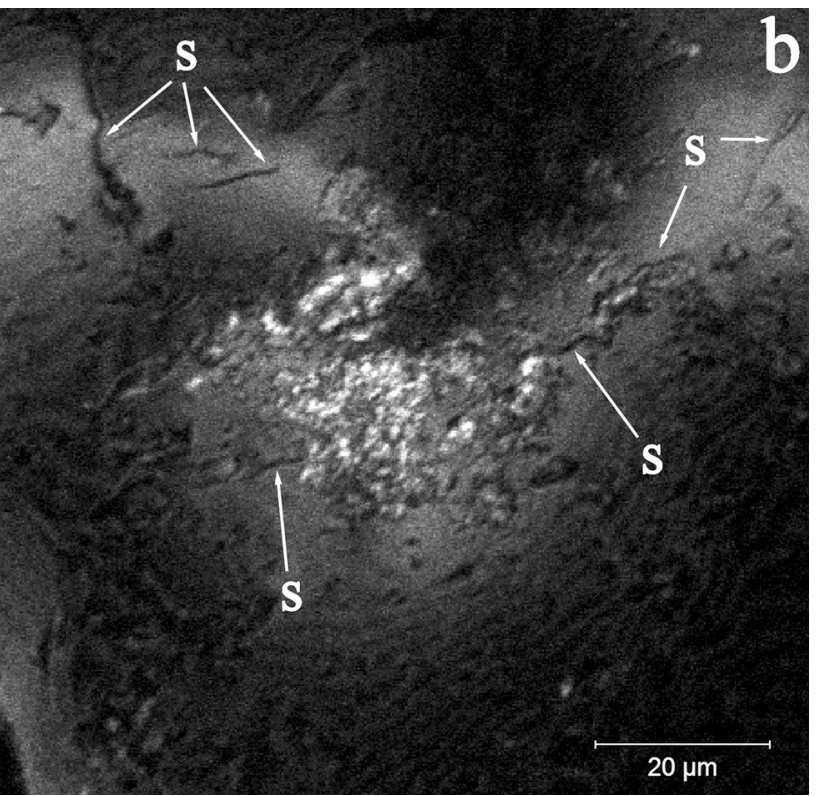}\\~\\
\includegraphics[scale=.6]{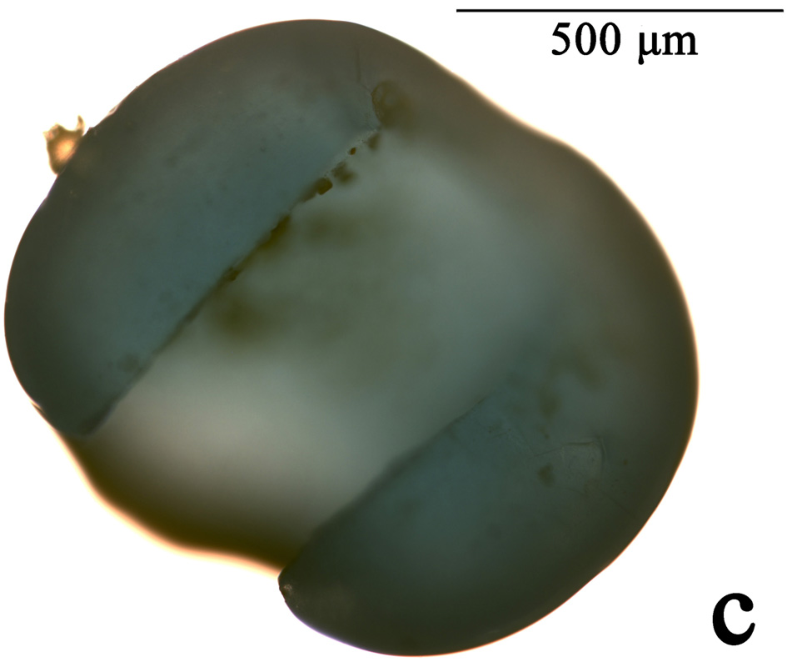}
\caption{\label{fig:lsm-710}(Color online) 
LSM images (a),\,(b) of the cleavage surface of a segmented bead shown in the panel (c); numerous foreign microinclusions are spread throughout the spall area and aggregation of the inclusions is observed in the centres of the images (a) and (b); cracks of various sizes are also observed everywhere on the surface (some of them are marked by the letters `S') which indicate the  presence of a strong strain.
}
\end{figure*}

Numerous microinclusions and microcracks of various sizes from tens nanometers to several micrometers are observed on the cleavage surface of a segmented bead shown in Fig.\,\ref{fig:lsm-710}. In addition, a rather large aggregation of some clusters is also seen   in the same images. We believe that these inhomogeneities are the best candidates to the role of the sought-for scatterers of light.

Let us dwell on possible reasons of bead cracking. To explain this phenomenon one should look for a source of an internal stress resulting in local inelastic deformation and rupture of glass. We suppose that some big enough precipitates, likely crystalline,  might be such sources.  They seem to be available in the bead glass since LSM images of some inclusions in Fig.\,\ref{fig:lsm-710} resemble typical images faceted crystallites. They may arise as a result of decomposition, diffusion and crystallisation of some chemical components of glass, such as pigments, opacifiers, fluxes or stabilizers, during glass beads manufacturing. Emerging they give rise to a strain field resulting in both glass cracking and gettering of additional impurities and chemical components dissolved in glass which in turn form a new generation of microprecipitates and impurity atmospheres around the cracks and promote further internal corrosion of beads. 

From our viewpoint, this is a  probable model describing the process of destruction of the blue glass beads. For example, diffusion of copper  could increase its concentration in form of  CuO around cracks to values much exceeding 2\% and colour these regions in green. At the same time, antimony and its chemical compounds which were often utilized as opacifiers until mid 19th century \cite{EuroBeads} could form crystallites resembling those seen in Fig.\,\ref{fig:lsm-710}\,a,\,b (or, e.\,g., rhombohedral modification of Pb$_2$Sb$_2$O$_7$ observed in opaque yellow Roman glasses \cite{Roman_Glass:Pb2Sb2O7}); such precipitates might be sources of local strain required for cracking.  In addition, antimony oxide dyes glass in yellow that also might explain the observed changes of the beads colour in green. Moreover, formation of Sb-reach precipitates and/or migration of Sb to strained domains reduces its concentration in the form of Sb$_2$O$_3$ dissolved in glass as consequence increasing the fraction of  ferrous oxide (Fe$^{2+}$) and dyeing glass in green.

However, presently this model is only a hypothesis which, to be confirmed or refuted, requires further explorations. Moreover, the proposed physical mechanism of bead degradation does not exclude a possibility of chemical corrosion, both on the surface and in the bulk of beads. Processes of chemical surface corrosion mentioned in Section\,\ref{sec:intro} may accompany the proposed physical processes of the bulk corrosion. Additionally, fracturing of beads resulting in formation of a developed volumetric network of cracks reaching the glass surface may stimulate penetration of chemical agents, say water and/or dissolved ions, into the bead bulk. In this case chemical corrosion may start in the glass bulk which facilitates crumbling of beads.

\section{\label{sec:conclusion}Conclusion}

Summarizing the above we would like to emphasise the main statements of the paper.

We have analysed  chemical composition of the blue-green beads by means of the XRF spectroscopy and found that they had been  made of lead-potassium glass with rather high content of copper; appreciable quantities of calcium, arsenic, antimony and iron are also present in the glass.

We have studied a  microstucture of the blue-green beads and explored its evolution in the process of bead deterioration. 
Investigated transmittance and scattering spectra in visible and near infrared ranges we have detected emergence of microscopic inhomogeneities with the sizes much less than 150\,nm inside the beads and significant growth of their density during the bead degradation. 
By means of laser scanning microscopy we have also observed numerous microscopic foreign inclusions and fine rifts on the cleavage surface of a fragmented bead.

We have come to conclusion that the most probable physical driving force of destruction of the blue-green beads is the strain induced by the crystalline microinclusions precipitated during and/or after the bead manufacturing and by their agglomerates which gives rise to internal microcracks and extended impurity atmospheres; the latter arise due to strain-stimulated diffusion of some  components of glass, such as potassium, copper and antimony, into the domains with high density of the precipitated clusters denuding the rest volume of the beads and introducing chemical inhomogeneity into the glass. Bead granulation at the last phase of corrosion seems to result from this inhomogeneity. The precipitates responsible for the internal strain in the glass presumably consist of some chemical compound of antimony.

Concluding the article we would like to notice that in the past decades optical techniques based on Rayleigh scattering of light were often used for characterization and imaging  of microdefects in various materials such as single-crystalline semiconductors \cite{Ogawa_RSI, Yuryev_RSI, Yuryev_SEM&SC,Sakai_RSI} or leucosapphire wafers \cite{Sapphire_Astf}, etc. This article demonstrates that such methods are informative for investigation of historical or archaeological glasses and glassware usually subjected to different kinds of volumetric and surface corrosion.

\ack
The research was carried out under the Collaboration Agreement between the State Research Institute for Restoration and A.\,M.\,Prokhorov General Physics Institute of RAS. 
The Center of Collective Use of Scientific Equipment of GPI RAS supported this research by presenting admittance to its instrumentation.
We cordially thank Prof. Victor B. Loschenov and especially Dr. Anastasia V. Ryabova for examining the samples of beads by means of the laser scanning microscope. We express our appreciation to Mr. Ilya B. Afanasyev of the Forensic Science Center of the Ministry of the Interior of the Russian Federation for the XRF measurements and the elemental analysis. We also thank Mr. Nikolay Yuryev for his help in making and assembling the microspectrometer.

\section*{References}

\bibliography{Green-blue_beads---light_transmission}

\end{document}